# Comment on Nongeometric Conditional Phase Shift via Adiabatic Evolution of Dark Eigenstate

Hua Zhong Li*

Advanced Research Center, Sun Yat-sen (Zhongshan) University, Guangzhou 510275, China

In a recent letter [1], it is claimed that based on a new kind of quantum mechanical phase of wave function which is neither dynamical nor geometrical a new kind of phase gate for quantum computation is discovered. The origin of this phase is the adiabatic evolution of a dark state. I would like to point out that these claims are questionable. The so called new phase gate was based on a phase which was nothing new but has been known for decades. It is something which has no physical meaning and is not observable. It was also pointed out by [4] that [1] was incorrect and the phase is the known geometric phase. Here we present a totally different argument.

(1) It is well-known that for a physical system with time-dependent Hamiltonian H(t), energy of the system is not conserved. It is also known that for a physical system having three or more energy levels correlated with time-dependent Hamiltonian there exist a very special evolution called "adiabatic passage" or "adiabatic following" in which the system follows dark state with zero eigenvalue [2]. The idea of using dark state to form a phase gate was already proposed in [3]. In this case of [1], dark states levels must be degenerate. In [1] the $D_1$ and $D_1'$ levels are degenerate. As generally known, the phases for degenerate levels is a matrix not a single number. During the evolution of the dark states in [1] the intermediates states or the final state are linear superposition of the degenerate states. [1] did not show how to define his "new" phase for the evolution along a open path. For a cyclic Hamiltonian, the wave function of dark states acquires a geometric phase and no dynamical phase. The geometric phase appears only after a cycle of evolution. If the system does not complete a cycle, phase is not observable and can be eliminated by a phase transformation. In the article[1] it is assumed that the evolution is adiabatic and the system does not follow a closed loop path in parameter space and at the same-time it requires the system to keep in a dark state at every instant with instantaneous eigenvalue of instantaneous Hamiltonian of H(t), E(t)=0 always. This is contradictive and impossible. Since in [1] the Hamiltonian H(t) expressed as formula (1) in [1] was parameterized by $l_1, l_2$ and $l_3$. H(t) was separately defined by different $l_3$, in two stages, one by $l_3$ another one by $-l_3$. The dark states $D_1$ and $D_1'$ (formular (2) and (5) in [1]) are also expressed in terms of $l_1, l_2$, and $l_3$. So the dark state eigenfunction in two stages are not a single continuous function. They do not form a single evolution solution of H(t) at two different stages. They are two different solutions of two different

*Email address: puslhz@mail.sysu.edu.cn

Hamiltonians at two stages without stage boundary condition specified. They can only be solved with two Schroedinger equations. The parameter $l_3$ from stage 1 to stage 2 experiences a discontinuity. The dark states in each stage are doubly degenerate. The final state of stage 1 is a superposition of dark states and become the initial state of stage 2. The dark state wave function for the second stage are not in general the same as the first stage due to different initial conditions. In [1] it is assumed that these two stages give an open path phase of each stage and claimed that in this process there exists a phase in the wave function which is neither geometric nor dynamic. In the next section of the present note, we show that even in the nondegenerate case along the open path evolution the so called nondynamic and non-geometric phase is unobservable and can be eliminated by phase transform.

(2) Let us discuss the formal solution of Schroedinger equation with a time-dependent Hamiltionan. For simplicity we discuss a nondegenerate level. Analogous procedure can be generalized to degenerate states. It is well-known that the system can be solved formally and exactly by Lewis-Riesenfeld's invariant method [5]. It was known that for adiabatic evolution [7] the solution can be expressed by

$$y(\vec{x}) = e^{ia(t)} |n(t)\rangle$$

where $|n(t)\rangle$ is an eigenfunction of an invariance operator with respect to H(t) as defined in [5], and

$$a(t) = d(t) + g(t) \tag{1}$$

Where $a(t)$ is the total phase, $d(t)$ can be defined as the usual dynamical phase and a general form of $g(t)$ is [5,7]

$$g(t) = -\int_0^t \langle n(t') | \frac{\partial}{\partial t} | n(t') \rangle dt' \tag{2}$$

$$l_g = -\oint \langle n(t') | \frac{\partial}{\partial t} | n(t') \rangle dt' \tag{3}$$

If the evolution is adiabatic non-cyclic as assumed by [1], it is well-known that this $g(t)$ can be eliminated with a phase transformation without resulting any physical effect. This last point was already known in 50 years ago and was written in a very famous textbook of quantum mechanics by L. I. Schiff [6]. This type of phase $g(t)$ is not the dynamic phase and not geometric phase in general. [1] anounced a phase which is neither controlled by the Hamiltonian and nor geometric is just of this type not a new phase. The formal solution of Schroedinger equation by invariant method is general and excludes the other phases besides the above.

The phase claimed in [1] does not give physical observable effect and is not able to operate as a phase gate in quantum computation.

Besides the claim of discovering a new kind of phase gate, the idea of this paper[1] of applying dark state to eliminate dynamic phase was already given by Duan *et al.* [3] and although this paper was cited in reference of [1] but only as an example of ion trap method and no due credit was given.